\documentclass[10pt,conference]{IEEEtran}
\IEEEoverridecommandlockouts
\usepackage{cite}
\usepackage{amsmath,amssymb,amsfonts}
\usepackage{algorithmic}
\usepackage{graphicx}
\usepackage{textcomp}
\usepackage{xcolor}
\usepackage{units}
\usepackage{balance}

\usepackage{acro}

\DeclareAcronym{5g}{
short=5G,
long= fifth generation,
}
\DeclareAcronym{6g}{
short=6G,
long= sixth generation,
}
\DeclareAcronym{3d}{
short=3D,
long= three-dimensional,
}
\DeclareAcronym{2d}{
short=2D,
long= two-dimensional,
}
\DeclareAcronym{aod}{
short=AOD,
long= angle-of-departure,
}
\DeclareAcronym{aosa}{
short=AOSA,
long= array-of-subarray,
}
\DeclareAcronym{adod}{
short=ADOD,
long= angle-difference-of-departure,
}
\DeclareAcronym{aoa}{
short=AOA,
long= angle-of-arrival,
}
\DeclareAcronym{adc}{
short=ADC,
long= analog to digital converter,
}
\DeclareAcronym{aeb}{
short=AEB,
long= angle error bound,
}
\DeclareAcronym{av}{
short=AV,
long= autonomous vehicle,
}
\DeclareAcronym{bs}{
short=BS,
long= base station,
}

\DeclareAcronym{csi}{
short=CSI,
long= channel state information,
}
\DeclareAcronym{cfo}{
short=CFO,
long= carrier frequency offset,
}
\DeclareAcronym{ceb}{
short=CEB,
long= clock error bound,
}
\DeclareAcronym{coa}{
short=COA,
long= curvature-of-arrival,
}
\DeclareAcronym{crb}{
short=CRB,
long= Cram\'er-Rao bound,
}
\DeclareAcronym{ccrb}{
short=CCRB,
long= constrained Cram\'er-Rao bound,
}
\DeclareAcronym{cmos}{
short=CMOS,
long= complementary metal-oxide-semiconductor,
}
\DeclareAcronym{crlb}{
short=CRLB,
long= Cram\'er-Rao lower bound,
}
\DeclareAcronym{cdf}{
short=CDF,
long= cumulative distribution function,
}
\DeclareAcronym{cp}{
short=CP,
long= cyclic prefix,
}
\DeclareAcronym{dac}{
short=DAC,
long= digital to analog converter,
}
\DeclareAcronym{dfl}{
short=DFL,
long= device-free localization,
}
\DeclareAcronym{dmimo}{
short=D-MIMO,
long= distributed MIMO,
}
\DeclareAcronym{dlprs}{
short=DL-PRS,
long= downlink positioning reference signal,
}
\DeclareAcronym{d2d}{
short=D2D,
long= device-to-device,
}
\DeclareAcronym{dftsofdm}{
short=DFT-s-OFDM,
long= discrete-Fourier-transform spread OFDM,
}
\DeclareAcronym{dl}{
short=DL,
long= deep learning,
}
\DeclareAcronym{gps}{
short=GPS,
long= global positioning system,
}
\DeclareAcronym{gnss}{
short=GNSS,
long= global navigation satellite system,
}
\DeclareAcronym{fim}{
short=FIM,
long = Fisher information matrix,
}
\DeclareAcronym{hwi}{
short=HWI,
long= hardware impairment,
}
\DeclareAcronym{hemt}{
short=HEMT,
long= high electron mobility transistor,
}
\DeclareAcronym{hbt}{
short=HBT,
long= heterojunction bipolar transistors,
}
\DeclareAcronym{iot}{
short=IoT,
long= internet of things,
}
\DeclareAcronym{imu}{
short=IMU,
long= inertial measurement unit,
}
\DeclareAcronym{isac}{
short=ISAC,
long= integrated sensing and communication,
}
\DeclareAcronym{iqi}{
short=IQI,
long= in-phase and quadrature imbalance,
}
\DeclareAcronym{ip}{
short=IP,
long= incidence point,
}
\DeclareAcronym{ia}{
short=IA,
long= initial access,
}
\DeclareAcronym{kpi}{
short=KPI,
long= key performance indicator,
}
\DeclareAcronym{kf}{
short=KF,
long= Kalman filter,
}
\DeclareAcronym{ekf}{
short=EKF,
long= extended Kalman filter,
}
\DeclareAcronym{ukf}{
short=UKF,
long= unscented Kalman filter,
}
\DeclareAcronym{ckf}{
short=CKF,
long= cubature Kalman filter,
}
\DeclareAcronym{pf}{
short=PF,
long= particle filter,
}
\DeclareAcronym{lb}{
short=LB,
long= lower bound,
}
\DeclareAcronym{lse}{
short=LSE,
long= least-square estimator,
}
\DeclareAcronym{lo}{
short=LO,
long= local oscillator,
}
\DeclareAcronym{los}{
short=LOS,
long= line-of-sight,
}
\DeclareAcronym{mc}{
short=MC,
long= mutual coupling,
}
\DeclareAcronym{mac}{
short=MAC,
long= medium access control,
}
\DeclareAcronym{meb}{
short=MEB,
long= mapping error bound,
}
\DeclareAcronym{ml}{
short=ML,
long= machine learning,
}
\DeclareAcronym{mcrb}{
short=MCRB,
long= misspecified Cram\'er-Rao bound,
}
\DeclareAcronym{mds}{
short=MDS,
long= multidimensional scaling ,
}
\DeclareAcronym{mimo}{
short=MIMO,
long= multiple-input-multiple-output,
}
\DeclareAcronym{siso}{
short=SISO,
long= single-input-single-output,
}
\DeclareAcronym{mm}{
short=MM,
long= mismatched model,
}
\DeclareAcronym{mpc}{
short=MPC,
long= multipath components,
}
\DeclareAcronym{mmwave}{
short=mmWave,
long= millimeter wave,
}
\DeclareAcronym{mmle}{
short=MMLE,
long= mismatched maximum likelihood estimation,
}
\DeclareAcronym{mems}{
short=MEMS,
long= micro-electro-mechanical system,
}
\DeclareAcronym{mle}{
short=MLE,
long= maximum likelihood estimation,
}
\DeclareAcronym{nlos}{
short=NLOS,
long= none-line-of-sight,
}
\DeclareAcronym{ofdm}{
short=OFDM,
long= orthogonal frequency-division multiplexing,
}
\DeclareAcronym{oeb}{
short=OEB,
long= orientation error bound,
}
\DeclareAcronym{otfs}{
short=OTFS,
long= orthogonal time-frequency space,
}
\DeclareAcronym{pdf}{
short=PDF,
long= probability density function,
}
\DeclareAcronym{papr}{
short=PAPR,
long= peak-to-average-power ratio,
}
\DeclareAcronym{pan}{
short=PAN,
long= power amplifier nonlinearity,
}
\DeclareAcronym{pa}{
short=PA,
long= power amplifier,
}
\DeclareAcronym{ps}{
short=PS,
long= phase shifter,
}
\DeclareAcronym{pn}{
short=PN,
long= phase noise,
}
\DeclareAcronym{poa}{
short=POA,
long= phase-of-arrival,
}
\DeclareAcronym{pwm}{
short=PWM,
long= planar wave model,
}
\DeclareAcronym{pdoa}{
short=PDOA,
long= phase-difference-of-arrival,
}
\DeclareAcronym{prs}{
short=PRS,
long= positioning reference signals,
}
\DeclareAcronym{peb}{
short=PEB,
long= position error bound,
}
\DeclareAcronym{rnn}{
short=RNN,
long= recurrent neural network,
}
\DeclareAcronym{rl}{
short=RL,
long= reinforcement learning,
}
\DeclareAcronym{rfc}{
short=RFC,
long= radio-frequency chain,
}
\DeclareAcronym{rf}{
short=RF,
long= radio frequency,
}
\DeclareAcronym{rfid}{
short=RFID,
long= radio frequency identification,
}
\DeclareAcronym{ris}{
short=RIS,
long= reconfigurable intelligent surface,
}
\DeclareAcronym{rss}{
short=RSS,
long= received signal strength,
}
\DeclareAcronym{rtt}{
short=RTT,
long= round-trip time,
}
\DeclareAcronym{sm}{
short=SM,
long= standard model,
}
\DeclareAcronym{sige}{
short=SiGe,
long= silicon-germanium,
}
\DeclareAcronym{spp}{
short=SPP,
long= surface plasmon polariton,
}
\DeclareAcronym{sa}{
short=SA,
long= subarray,
}
\DeclareAcronym{sota}{
short=SOTA,
long= state-of-the-art,
}
\DeclareAcronym{swm}{
short=SWM,
long= spherical wave model,
}
\DeclareAcronym{slam}{
short=SLAM,
long= simultaneous localization and mapping,
}
\DeclareAcronym{tm}{
short=TM,
long= true model,
}
\DeclareAcronym{toa}{
short=TOA,
long= time-of-arrival,
}
\DeclareAcronym{tof}{
short=TOF,
long= time-of-flight,
}
\DeclareAcronym{tdoa}{
short=TDOA,
long= time-difference-of-arrival,
}
\DeclareAcronym{thz}{
short=THz,
long= Terahertz,
}
\DeclareAcronym{ue}{
short=UE,
long= user equipment,
}
\DeclareAcronym{ummimo}{
short=UM-MIMO,
long= ultra-massive multi-input-multi-output,
}
\DeclareAcronym{vlp}{
short=VLP,
long= visible light positioning,
}
\DeclareAcronym{veb}{
short=VEB,
long= velocity error bound,
}
\DeclareAcronym{vlc}{
short=VLC,
long= visible light communication,
}
\DeclareAcronym{ula}{
short=ULA,
long= uniform linear array,
}
\DeclareAcronym{uav}{
short=UAV,
long= unmanned aerial vehicle,
}
\DeclareAcronym{upa}{
short=UPA,
long= uniform planar array,
}
\DeclareAcronym{wlan}{
short=WLAN,
long= wireless local area network,
}

\usepackage{tabu,longtable}
\usepackage{color}

\usepackage{tikz} 
\usepackage[utf8]{inputenc}
\usepackage{pgfplots} 
\usepackage{pgfgantt}
\usepackage{pdflscape}
\usepackage{bm}

\def\BibTeX{{\rm B\kern-.05em{\sc i\kern-.025em b}\kern-.08em
    T\kern-.1667em\lower.7ex\hbox{E}\kern-.125emX}}
    
\IEEEoverridecommandlockouts\IEEEpubid{\makebox[\columnwidth]{ 978-1-6654-3540-6/22~\copyright~2022 IEEE \hfill} \hspace{\columnsep}\makebox[\columnwidth]{ }}

\begin{document}
\pdfoutput=1

\title{Localization Coverage Analysis of THz Communication Systems with a 3D Array}
\author{
Pinjun Zheng\IEEEauthorrefmark{1}, 
Tarig Ballal\IEEEauthorrefmark{1}, 
Hui Chen\IEEEauthorrefmark{2}, 
Henk Wymeersch\IEEEauthorrefmark{2}, 
Tareq Y. Al-Naffouri\IEEEauthorrefmark{1}\\
\IEEEauthorrefmark{1}King Abdullah University of Science and Technology, Thuwal, KSA\\
\IEEEauthorrefmark{2}Chalmers University of Technology, Gothenburg, Sweden\\
E-mail: pinjun.zheng@kaust.edu.sa
}


\maketitle

\begin{abstract}
This paper considers the problem of estimating the position and orientation of a user equipped with a three-dimensional (3D) array receiving downlink far-field THz signals from multiple base stations with known positions and orientations. We derive the Cram\'{e}r–Rao Bound for the localization problem and define the coverage of the considered system.
We compare the error lower bound distributions of the conventional planar array and the 3D array configurations at different \ac{ue} positions and orientations. Our numerical results obtained for array configurations with an equal number of elements show very limited coverage of the planar-array configuration, especially across the UE orientation range. Conversely, a 3D array configuration offers an overall higher coverage with minor performance loss in certain UE positions and orientations.
\end{abstract}


\section{Introduction}
With the increasing demands for higher data traffic in wireless communication, \ac{thz} frequency band ($\unit[0.1\text{-}10]{THz}$) is envisioned as a key enabler for future \ac{6g} wireless communication systems and beyond~\cite{9112745,sarieddeen2021overview}.
In addition to the benefits to communication, larger array size (high angular resolution) and larger bandwidth (high delay resolution) in high-frequency systems also enable high-accuracy localization, which has been extensively explored within \ac{mimo} communication systems~\cite{8240645,9721709}.
It is foreseeable that the potential distance-/angle-aware applications, such as virtual reality (VR)/augmented reality (AR)~\cite{saeedi2018navigating}, vehicular safety~\cite{9665433}, \ac{gnss}~\cite{liu2021constrained}, etc, will be further exploited in the future communication systems~\cite{9112745,chen2021tutorial}.

In geometry-based localization, the position and orientation information is usually estimated from the geometrical measurements such as \ac{toa}, \ac{tdoa}, \ac{aod}, \ac{aoa}, etc~\cite{chen2021tutorial}.
However, techniques based on the time/range related measurements like TOA and TDOA require tight synchronization. 
To avoid this, the angle-based localization methods were pursued~\cite{nazari20213d}, which require the receiver/transmitter to be equipped with antenna arrays.
Over the years, a plethora of localization techniques based on 1D uniform arrays~\cite{6784422,9086740} and 2D uniform arrays~\cite{6850012,8058460} have been proposed. Localization methods using arbitrary array configurations have also been proposed, e.g.,~\cite{6638438}.
Moreover, a few works demonstrate the application of nonuniform array design for purposes such as short-range localization~\cite{8474363}, 
joint design with signal detection~\cite{7740042}, cost reduction~\cite{8058460}, etc.

Although promising localization results are shown in these works and some localization-aware mobile network deployment solutions are proposed~\cite{albanese2022loko}, the coverage issue is rarely discussed, limiting the availability of providing localization services. 
To improve the coverage and enhance the connectivity, 3D arrays could be one of the promising techniques.
In~\cite{7096364}, a localization error bound analysis of 2D and 3D V-shaped arrays is reported.
Despite the potential of 3D arrays in localization applications, this concept has not yet been widely studied in radio localization due to an impractical physical size. However, we expect this issue can be solved in the THz band with a much smaller signal wavelength\footnote{For example, a $10\times 10$ half-wavelength spaced array of a $\unit[140]{GHz}$ system can be fitted into a $\unit[1]{cm^2}$ area, while the same footprint can only support a $2\times 2$ array at the frequency of $\unit[28]{GHz}$.}.
Besides, since the existing works on 3D array localization are limited to the specific array configurations such as V-shaped arrays in~\cite{7096364}, a more general model for 3D array localization is expected.

In this paper, we consider a downlink, far-field THz band MIMO scenario where the angles and delay measurements from multiple \acp{bs} are used to estimate the position and orientation of a \ac{ue}. In THz communication, a commonly used array configuration is the array-of-subarrays (AoSA) structure, which can mitigate the high-frequency hardware constraints and support low-complexity beamforming~\cite{lin2016terahertz,9591285}. In this work, we consider arranging the \acp{sa} of a UE in the 3D space with arbitrary known positions and orientations, whereas each \ac{sa} is arranged in a 2D space (i.e., a planar \ac{sa}). Our investigation reveals that deploying such 3D array configurations for THz localization can improve coverage relative to the conventional 2D (or planar) structures.
The main contributions of this paper are as follows:
\begin{itemize}
    \item We derive the \ac{peb} and \ac{oeb} for the underlying localization problem in the form of the \ac{ccrb}.
    \item We provide a comparative performance analysis of a 2D and a 3D array using the PEB and OEB distribution across different UE positions and orientations. 
    \item We define the localization coverage using the cumulative distribution function (CDF) of the PEB and OEB, and give a quantitative evaluation of the coverage for 2D and 3D array configurations. 
\end{itemize}

\section{Problem Statement}

\subsection{System Model}

\begin{figure}[t]
    \centering
    \includegraphics[width=3.5in]{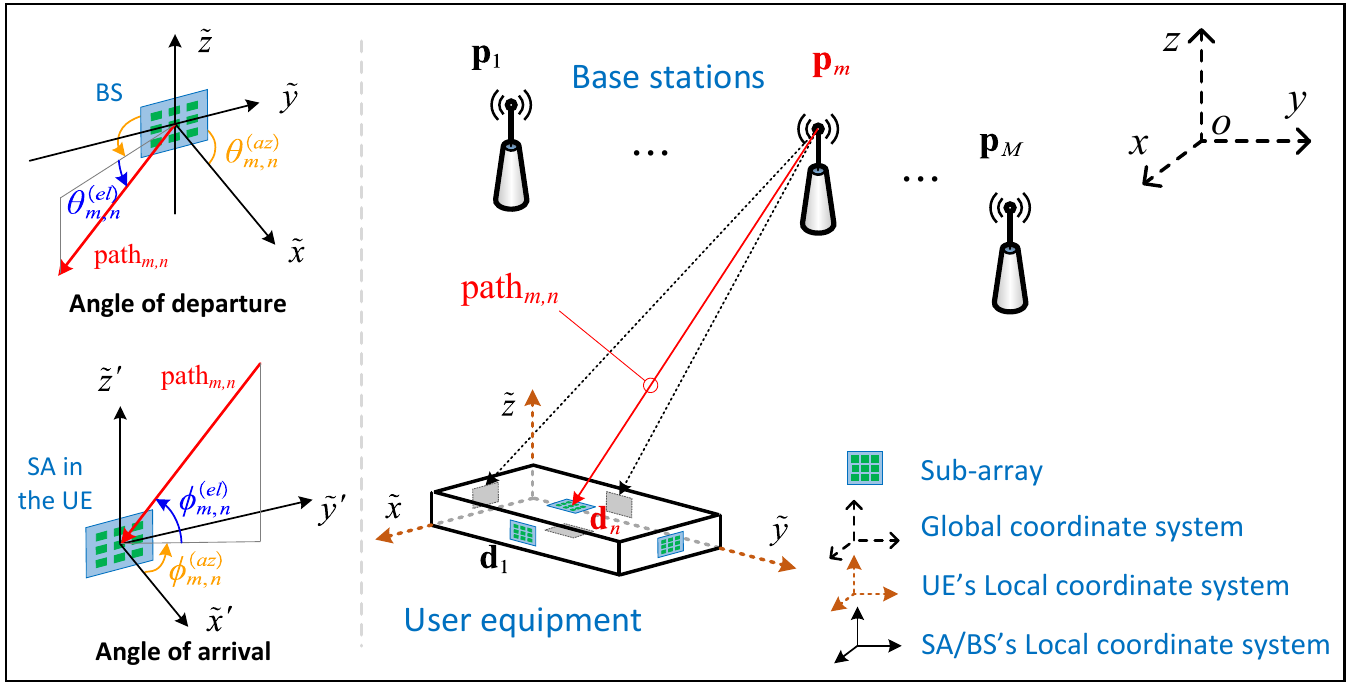}
    \caption{Illustration of a localization scenario in a communication system with multiple BSs and 1 UE with multiple SAs. The left part shows the geometry of the azimuth and elevation components of AOD and AOA measurements.}
    \label{fig_1}
  \end{figure}

We consider a far-field downlink scenario with $M$ \acp{bs} and $1$ UE as shown in Fig.~\ref{fig_1}. 
The positions and orientations of the BSs are known in a global coordinate system. Each BS is equipped with a planar array 
to provide AOD measurements. 
The UE consists of $N$ SAs arranged in a 3D space with fixed relative positions and orientation, and each SA provides  AOA measurements with respect to each BS if the \ac{los} channel exists. 

We use $\{\mathbf{p}_{\text{B},m}\in\mathbb{R}^{3\times 1} \}_{m=1}^M$ to denote the BS positions in the global coordinate system, and the rotation matrix $\{\mathbf{R}_{\text{B},m}\in\mathbb{R}^{3\times 3}\}_{m=1}^M$ to represent the BS orientations. The position and orientation of UE that are to be estimated are denoted as $\mathbf{p}_\text{U}\in\mathbb{R}^{3\times 1}$ and $\mathbf{R}_\text{U}\in\mathbb{R}^{3\times 3}$, respectively.
Moreover, we use $\{\mathbf{d}_n\in\mathbb{R}^{3\times 1}\}_{n=1}^N$ and $\{\mathbf{R}_n\in\mathbb{R}^{3\times 3}\}_{n=1}^N$ to denote the known \emph{relative} positions and orientations of the $N$ SAs in UE's local coordinate system, respectively. All the rotation matrix lies in the orthogonal group SO(3)~\cite{nazari20213d} that satisfy the constraints
\begin{equation}\label{eq_cons}
    \begin{cases}
        \mathbf{R}^\mathsf{T}\mathbf{R}=\mathbf{I},\\
        \det(\mathbf{R}) = 1.
    \end{cases} 
\end{equation}

The rotation matrix represents the rotation relationship between a global and a local coordinate system. For example, given a vector $\mathbf{a}$ in the global coordinate system, we can obtain its coordinates in UE's local coordinate system as
$
    \widetilde{\mathbf{a}} = \mathbf{R}_{\text{U}}^\mathsf{T} \mathbf{a}.
$
By using rotation matrices, we can also express the positions and orientations of SAs in the global coordinate system. The position of the center of the $n$-th SA is given by 
$
  \mathbf{p}_n = \mathbf{p}_{\text{U}} + \mathbf{R}_{\text{U}}\mathbf{d}_n.
$
For orientation, 
the coordinates of $\mathbf{a}$ in the local coordinate system of the $n$-th subarray is 
$
  \widetilde{\mathbf{a}}_n = \mathbf{R}_n^\mathsf{T} \mathbf{R}_{\text{U}}^\mathsf{T} \mathbf{a}.
$
\subsection{Signal Model}\label{sec_1}

In THz band, the non-line-of-sight (NLOS) paths become increasingly sparse and lossy~\cite{chen2021tutorial}, so we consider the LOS path only.
Consider the far-field OFDM MIMO channel model~\cite{9591285}, the received signal at 
the $k$-th subcarrier and the $g$-th transmission in the path from BS $m$ to SA $n$ in the UE is
\begin{equation}\label{channel}
  {y}_{(m,n)}^{(g)}[k] = \underbrace{\sqrt{P}\mathbf{w}_{\text{U}}^\mathsf{T} \mathbf{H}(k,\bm{\eta}_{(m,n)})\mathbf{w}_{\text{B}}{x}^{(g)}[k]}_{\bm{\mu}^{(g)}[k]} + \mathbf{w}_{\text{U}}^\mathsf{T}\mathbf{n}^{(g)}[k],
\end{equation}
where $P$ is the average transmission power, 
$\mathbf{w}_{\text{U}}\in \mathbb{C}^{N_\text{U} \times 1}$ is the combiner vector at the SA of the UE, 
$\mathbf{H}(k,\bm{\eta}_{(m,n)})\in\mathbb{C}^{N_\text{U} \times N_\text{B}}$ is the channel matrix~\cite{chen2021tutorial},
$\mathbf{w}_{\text{B}}\in \mathbb{C}^{N_B \times 1}$ is the precoder vector at the BS,
${x}^{(g)}[k]\in\mathbb{C}$ is the signal symbol vector before the precoder,
and $\mathbf{n}^{(g)}[k]\sim \mathcal{CN}(\mathbf{0}, \sigma_n^2\mathbf{I}_{N_U})$. Finally, $\bm{\eta}_{(m,n)}$ denotes the channel parameters from BS $m$ to SA $n$, i.e., $\bm{\eta}_{(m,n)}=[\theta_{m,n}^{(\text{az})},\theta_{m,n}^{(\text{el})},\phi_{m,n}^{(\text{az})},\phi_{m,n}^{(\text{el})},\tau_{m,n}]^\mathsf{T}$. The AOD pair consists of an azimuth angle $\theta_{m,n}^{(\text{az})}$ an elevation angle $\theta_{m,n}^{(\text{el})}$, 
while the AOA pair consists of an azimuth angle $\phi_{m,n}^{(\text{az})}$  and an elevation angle $\phi_{m,n}^{(\text{el})}$, as visualized in Fig.~\ref{fig_1}. The parameters are defined as 
\begin{align}\label{eq_forward1}
    \theta_{m,n}^{(\text{az})} & = \arctan2\Bigl(\mathbf{u}_2^\mathsf{T} \mathbf{R}_{\text{B},m}^\mathsf{T} (\mathbf{p}_{\text{U}}+\mathbf{R}_{\text{U}}\mathbf{d}_n-\mathbf{p}_{\text{B},m}), \notag \\
    & \quad \quad \quad \quad \quad \quad \quad \mathbf{u}_1^\mathsf{T} \mathbf{R}_{\text{B},m}^\mathsf{T} (\mathbf{p}_{\text{U}}+\mathbf{R}_{\text{U}}\mathbf{d}_n-\mathbf{p}_{\text{B},m}) \Bigr),\\
    \theta_{m,n}^{(\text{el})} & = \arcsin\left( \frac{\mathbf{u}_3^\mathsf{T} \mathbf{R}_{\text{B},m}^\mathsf{T} (\mathbf{p}_{\text{U}}+\mathbf{R}_{\text{U}}\mathbf{d}_n-\mathbf{p}_{\text{B},m})}{\|\mathbf{p}_{\text{U}}+\mathbf{R}_{\text{U}}\mathbf{d}_n-\mathbf{p}_{\text{B},m}\|_2}\right),\\
    \phi_{m,n}^{(\text{az})}& = \arctan2\Bigl(-\mathbf{u}_2^\mathsf{T} \mathbf{R}_n^\mathsf{T} \mathbf{R}_\text{U}^\mathsf{T} (\mathbf{p}_{\text{U}}+\mathbf{R}_{\text{U}}\mathbf{d}_n-\mathbf{p}_{\text{B},m}), \notag \\
    & \quad \quad\quad \quad \quad\  -\mathbf{u}_1^\mathsf{T} \mathbf{R}_n^\mathsf{T} \mathbf{R}_\text{U}^\mathsf{T} (\mathbf{p}_{\text{U}}+\mathbf{R}_{\text{U}}\mathbf{d}_n-\mathbf{p}_{\text{B},m}) \Bigr),\\
    \phi_{m,n}^{(\text{el})}& = \arcsin\left( -\frac{\mathbf{u}_3^\mathsf{T} \mathbf{R}_n^\mathsf{T} \mathbf{R}_\text{U}^\mathsf{T} (\mathbf{p}_{\text{U}}+\mathbf{R}_{\text{U}}\mathbf{d}_n-\mathbf{p}_{\text{B},m})}{\|\mathbf{p}_{\text{U}}+\mathbf{R}_{\text{U}}\mathbf{d}_n-\mathbf{p}_{\text{B},m}\|_2}\right),\\\label{eq_forward5}
    \tau_{m,n} & = \frac{\|\mathbf{p}_{\text{U}}+\mathbf{R}_{\text{U}}\mathbf{d}_n-\mathbf{p}_{\text{B},m}\|_2}{c} + \rho,
\end{align}
where $\mathbf{u}_1 = [1, 0, 0]^\mathsf{T}  $, $\mathbf{u}_2 = [0, 1, 0]^\mathsf{T}  $, $\mathbf{u}_3 = [0, 0, 1]^\mathsf{T}  $,
$\rho$ models the total clock bias between the BS and UE, and $c$ is the speed of light. 
We assume that all the BSs are synchronized with each other and the SAs of the UE share the same clock signal. Hence, $\rho$ is fixed for all paths. 

In this work, each BS/SA is connected to an independent \ac{rfc} with a random precoder/combiner (phase-shifters have a constant amplitude with phases uniformly distributed from $0$ to $2\pi$) adopted for each transmission. In addition, orthogonal subcarriers are assigned to different BS-SA pairs to avoid the interference between different channels.
When the SAs in the UE receive a signal, we first estimate channel geometry parameters such as AOA, AOD, delay and complex gain, and then perform localization based on theses estimations.

\subsection{The Localization Problem} \label{sec:locproblem}

Considering a system with $M$ BSs and $1$ UE with $N$ SAs, we have at most $M\times N$ paths, each with an associated AOD, AOA, and delay.
However, depending on the position and orientation of the corresponding BS and SA, some paths may not be visible because of the limited radiation pattern of the antenna.
For example, if we assume all the antennas have a semi-sphere radiation pattern, the visibility of the paths can be modeled by defining a set of index pairs as
\begin{multline}\label{eq_Q}
    \mathcal{Q}:=\{(m,n)|\langle\mathbf{R}_{n}^\mathsf{T}\mathbf{R}_{\text{U}}^\mathsf{T}(\mathbf{p}_{\text{B},m}-\mathbf{p}_n),\tilde{\mathbf{e}}_n \rangle > 0,\\ \langle\mathbf{R}_{\text{B}}^\mathsf{T}(\mathbf{p}_n-\mathbf{p}_{\text{B},m}),\tilde{\mathbf{e}}_{\text{B},m} \rangle > 0\}, 
\end{multline} 
where $\langle\cdot,\cdot\rangle$ denotes the inner product, $\tilde{\mathbf{e}}_n$ and $\tilde{\mathbf{e}}_{\text{B},m}$ are respectively the normal vector of the SA $n$ and the normal vector of the array of the BS $m$ in their local coordinate systems, and the set $\mathcal{Q}$ contains the indexes of all the available paths.
To facilitate subsequent discussions, we assume the cardinality of the set $\mathcal{Q}$ is $D$ and assign new labels to the elements in the set $\mathcal{Q}$ from $1$ to $D$ as $\mathcal{Q}=\{(m_1,n_1), (m_2,n_2), \dots, (m_D,n_D)\}$.
The objective of the localization problem is to estimate the position $\mathbf{p}_{\text{U}}$ and orientation $\mathbf{R}_{\text{U}}$ of the UE 
from the available paths $\{\theta_{m,n}^{(\text{az})}, \theta_{m,n}^{(\text{el})}, \phi_{m,n}^{(\text{az})}, \phi_{m,n}^{(\text{el})}, \tau_{m,n}\}_{(m,n)\in \mathcal{Q}}$.

For each available path from BS $m$ to SA $n$ of the UE, we get AOD measurements on the BS side and AOA measurements on the UE side, as well as channel delay, from a channel estimator. 
We further stack all the available path parameters as 
\begin{equation}
\bm{\eta} = [\bm{\eta}_{(m_1,n_1)}^\mathsf{T},\ldots,\bm{\eta}_{(m_D,n_D)}^\mathsf{T}]^\mathsf{T}.
\end{equation}
Then the available measurement vector is $\hat{\bm{\eta}} \sim \mathcal{N}( \bm{\eta},\bm{\Sigma})$, where $\bm{\Sigma}$ is a block diagonal matrix, as the measurement vectors from different BSs are independent. 
%
%
The localization problem is then to determine $\mathbf{p}_\text{U}$ and $\mathbf{R}_\text{U}$, based on $\hat{\bm{\eta}}$.

\section{Error Bound and Performance Metrics}
We first derive the performance bound of the localization problem, from which the coverage metrics will be defined and analyzed.
The \ac{crb} is a useful tool as it gives a lower bound on the mean squared error (MSE). 
A brief introduction to \ac{crb} is given in the following subsection. More details can be found in, e.g.,~\cite{kay1993fundamentals,8356190}.

\subsection{Background on (Constrained) Cram\'{e}r–Rao Bound}
Consider the problem of estimating a deterministic unknown vector $\mathbf{x}\in\mathbb{R}^N$ from an observation $\mathbf{z}$ given a statistical model $p(\mathbf{z}|\mathbf{x})$.
The amount of information the observation carries about the unknown is measured by the Fisher information matrix (FIM), which is given by
\begin{equation}\label{eq_FIM1}
    \bm{\mathcal{I}} (\mathbf{x}) = \mathbb{E}_{\mathbf{z}}\{\nabla_{\mathbf{x}}\log p(\mathbf{z}|\mathbf{x})\nabla_{\mathbf{x}}^\mathsf{T} \log p(\mathbf{z}|\mathbf{x})\}.
\end{equation}
The FIM relates to the estimation error covariance of any unbiased estimator $\hat{\mathbf{x}}(\mathbf{z})$ as
\begin{equation}
    \mathbb{E}\{(\mathbf{x}-\hat{\mathbf{x}})(\mathbf{x}-\hat{\mathbf{x}})^\mathsf{T} \} \succeq \bm{\mathcal{I}}^{-1} (\mathbf{x}).
\end{equation}
Then a lower bound (which is known as \ac{crb}) on the estimation MSE is given by
\begin{equation}
    \mathbb{E}\{\|\mathbf{x}-\hat{\mathbf{x}}\|^2\} \geq \text{tr}\left(\bm{\mathcal{I}}^{-1} (\mathbf{x})\right).
\end{equation}


When the unknown vector $\mathbf{x}$ constrained to lie on a manifold $\mathbf{h}(\mathbf{x})=0$ defined by $0 \leq K< N$ non-redundant constraints, 
the error covariance is lower bounded by the \ac{ccrb} as~\cite{stoica1998cramer,nazari20213d}
\begin{equation}
    \mathbb{E}\{(\mathbf{x}-\hat{\mathbf{x}})(\mathbf{x}-\hat{\mathbf{x}})^\mathsf{T} \} \succeq \bm{\mathcal{I}}^{-1}_{\text{const}} (\mathbf{x}),
\end{equation}
where 
\begin{equation}\label{eq_constr}
    \bm{\mathcal{I}}_{\text{const}}^{-1} (\mathbf{x}) = \mathbf{M}\left(\mathbf{M}^\mathsf{T} \bm{\mathcal{I}} (\mathbf{x})\mathbf{M}\right)^{-1}\mathbf{M}^\mathsf{T},
\end{equation}
with $\bm{\mathcal{I}}_\text{const} (\mathbf{x})$ being the constrained FIM,
and $\mathbf{M}\in\mathbb{R}^{N\times(N-K)}$ satisfying
\begin{equation}\label{eq_M}
    \begin{cases}
        \mathbf{M}^\mathsf{T} \mathbf{M} = \mathbf{I}_{N-K},\\
        \frac{\partial \mathbf{h}(\mathbf{x})}{\partial \mathbf{x} }\cdot \mathbf{M} = \mathbf{0}_{K\times(N-K)},
    \end{cases} 
\end{equation}
and is obtained by collecting the orthonormal basis vectors of the null-space of the gradient matrix 
${\partial \mathbf{h}(\mathbf{x})}/{\partial \mathbf{x} }\in\mathbb{R}^{K\times N}$.

\subsection{Localization Performance Error Bound}
The FIM of the vector $\bm{\eta}_{(m,n)}$ can be obtained as~\cite{kay1993fundamentals}
\begin{align}
    &\bm{\mathcal{I}}({\bm{\eta}}_{(m,n)}) =\\
    & \frac{2}{\sigma^2}\sum_{g = 1}^{G} \sum_{k = 1}^{K}\text{Re}\left\{\left(\frac{\partial \bm{\mu}^{(g)}[k] }{\partial {\bm{\eta}}_{(m,n)}}\right)^\mathsf{H}
  \left(\frac{\partial \bm{\mu}^{(g)}[k] }{\partial {\bm{\eta}}_{(m,n)}}\right)\right\}.    \notag
\end{align}
Consequently, the FIM of all the channel parameters can be obtained as
\begin{equation}
\bm{\mathcal{I}}({\bm{\eta}})= \text{blkdiag}
\left\{\bm{\mathcal{I}}({\bm{\eta}}_{(m_1,n_1)}),\ldots,\bm{\mathcal{I}}({\bm{\eta}}_{(m_D,n_D)})\right\},
\end{equation}
where $\text{blkdiag}\{\cdot\}$ means forming a block diagonal matrix.

Now, considering the state vector $\mathbf{r} = [\mathbf{p}_\text{U}^\mathsf{T},\rho,\text{vec}(\mathbf{R}_\text{U})^\mathsf{T}]^\mathsf{T}$, we can see that $\bm{\eta}$ is a function of $\mathbf{r}$. These relationships are represented by the expressions in \eqref{eq_forward1}--\eqref{eq_forward5}.
Thus, the FIM with $\mathbf{r}$ as the estimation subject can be obtained as~\cite{nazari20213d}
\begin{equation}\label{eq_transformation}
    \bm{\mathcal{I} }(\mathbf{r}) = \mathbf{T}^\mathsf{T}\bm{\mathcal{I} }(\bm{\eta})\mathbf{T},
\end{equation}
where $[\mathbf{T}]_{i,j} = \partial \eta_i/\partial r_j$ are derived in Appendix \ref{appendix_1}.

Considering the constraints on $\mathbf{R}_\text{U}$ in \eqref{eq_cons} and according to \eqref{eq_constr}, we have 
\begin{equation}
    \bm{\mathcal{I} }_{\text{const}}^{-1}(\mathbf{r}) = \mathbf{M}(\mathbf{M}^\mathsf{T}\bm{\mathcal{I}}(\mathbf{r})\mathbf{M})^{-1}\mathbf{M}^\mathsf{T}.
\end{equation}
According to \eqref{eq_M}, a satisfied $\mathbf{M}$ can be 
\begin{equation}
  \mathbf{M} = 
\frac{1}{\sqrt{2}}\begin{bmatrix}
\sqrt{2}\mathbf{I}_{4\times 4} & \mathbf{0}_{4\times 1} & \mathbf{0}_{4\times 1} & \mathbf{0}_{4\times 1} \\
\mathbf{0}_{3\times 4} & -\mathbf{c}_3 & \mathbf{0}_{3\times 1} & \mathbf{c}_2 \\
\mathbf{0}_{3\times 4} & \mathbf{0}_{3\times 1} & -\mathbf{c}_3 & -\mathbf{c}_1 \\
\mathbf{0}_{3\times 4} & \mathbf{c}_1 & \mathbf{c}_2 & \mathbf{0}_{3\times 1} 
\end{bmatrix},
\end{equation}
where $ [\mathbf{c}_1,\mathbf{c}_2,\mathbf{c}_3] = \mathbf{R}_{\text{U}}$.
Therefore, we have the PEB and the OEB given by
\begin{align}
  \text{PEB} &= \sqrt{ \text{tr}([\bm{\mathcal{I} }^{-1}_{\text{const}}(\mathbf{r})]_{1:3,1:3})}, \\ 
  \text{OEB} &= \sqrt{ \text{tr}([\bm{\mathcal{I} }^{-1}_{\text{const}}(\mathbf{r})]_{5:13,5:13})}.
\end{align}

\subsection{Localization Coverage}

The coverage is a metric to evaluate the overall performance of a localization or communication system~\cite{liu2011localization,9271904}.
In this paper, We define the localization coverage as the probability that the PEB/OEB is lower than a threshold $\xi_p$/$\xi_o$ when the UE is at random positions $\mathbf{p}_\text{U}\in \Omega_p$ with  random orientations  $\mathbf{R}_\text{U}\in \Omega_R$ ($\Omega_p/\Omega_R$ are the space that UE position/orientation can be chosen). More specifically, the position coverage $C_p(\xi_p)$ and orientation coverage $C_o(\xi_o)$ can be defined as

\begin{equation}\label{eq_defcoverage}
\begin{aligned}
    C_p(\xi_p) &= \frac{\int_{\Omega_p}\int_{\Omega_o}H(\xi_p-\text{PEB}(\mathbf{p}_\text{U},\mathbf{R}_\text{U}))\text{d}\mathbf{p}_\text{U}\text{d}\mathbf{R}_\text{U}}
    {\int_{\Omega_p}\int_{\Omega_o}\text{d}\mathbf{p}_\text{U}\text{d}\mathbf{R}_\text{U}},\\
    C_o(\xi_o) &= \frac{\int_{\Omega_p}\int_{\Omega_o}H(\xi_p-\text{OEB}(\mathbf{p}_\text{U},\mathbf{R}_\text{U}))\text{d}\mathbf{p}_\text{U}\text{d}\mathbf{R}_\text{U}}
    {\int_{\Omega_p}\int_{\Omega_o}\text{d}\mathbf{p}_\text{U}\text{d}\mathbf{R}_\text{U}},
\end{aligned}
\end{equation} 
where 
$\text{d}\mathbf{p}_\text{U} = \text{d}{p}_{\text{U}x}\text{d}{p}_{\text{U}y}\text{d}{p}_{\text{U}z}$
, $\text{d} \mathbf{R}_\text{U} = \text{d}\alpha \text{d}\beta \text{d}\gamma$ \footnote{Since $\mathbf{R}_\text{U}$ has only three degrees of freedom~\cite{nazari20213d}, we denote them as $\{\alpha,\beta,\gamma\}$, which are the same as the Euler angles that will be introduced in subsection \ref{sec_simSce}.}, $\xi_p$ and $\xi_o$ are given thresholds for the position coverage and the orientation coverage respectively, and $H(\cdot)$  is the Heaviside step function (i.e., $H(t)=1, t\ge 0$ and zero elsewhere). 

Before presenting our simulation results, we can gain insight by analyzing the models. First, from the models in \eqref{eq_forward1}--\eqref{eq_forward5}, we can see that the UE orientation is related to the AODs 
$(\theta_{m,n}^{(\text{az})}$, $\theta_{m,n}^{(\text{az})})$ due to the term $\mathbf{R}_\text{U}\mathbf{d}_n$ only.
Again $\mathbf{d}_n$ is the position of sub-array $n$ in the UE's local coordinate system, whose scale is in the same order of magnitude as the scale of the size of the UE, much smaller than the distance between the UE and the BS. This means that rotating the UE (changing the orientation $\mathbf{R}_\text{U}$) does not cause much difference in term $\mathbf{p}_{\text{U}}+\mathbf{R}_{\text{U}}\mathbf{d}_n-\mathbf{p}_{\text{B},m}$, and thus not much difference in AODs. 
So, AODs carry limited information about the UE's orientation. On the contrary, AOAs carry most of the information regarding the UE rotation.
Therefore, it can be inferred that a lower OEB would appear in the cases with better AOA estimations. 


\section{Simulations}

\subsection{Simulation Scenario}\label{sec_simSce}
\begin{figure}[t]
    \centering
    \includegraphics[width=3.5in]{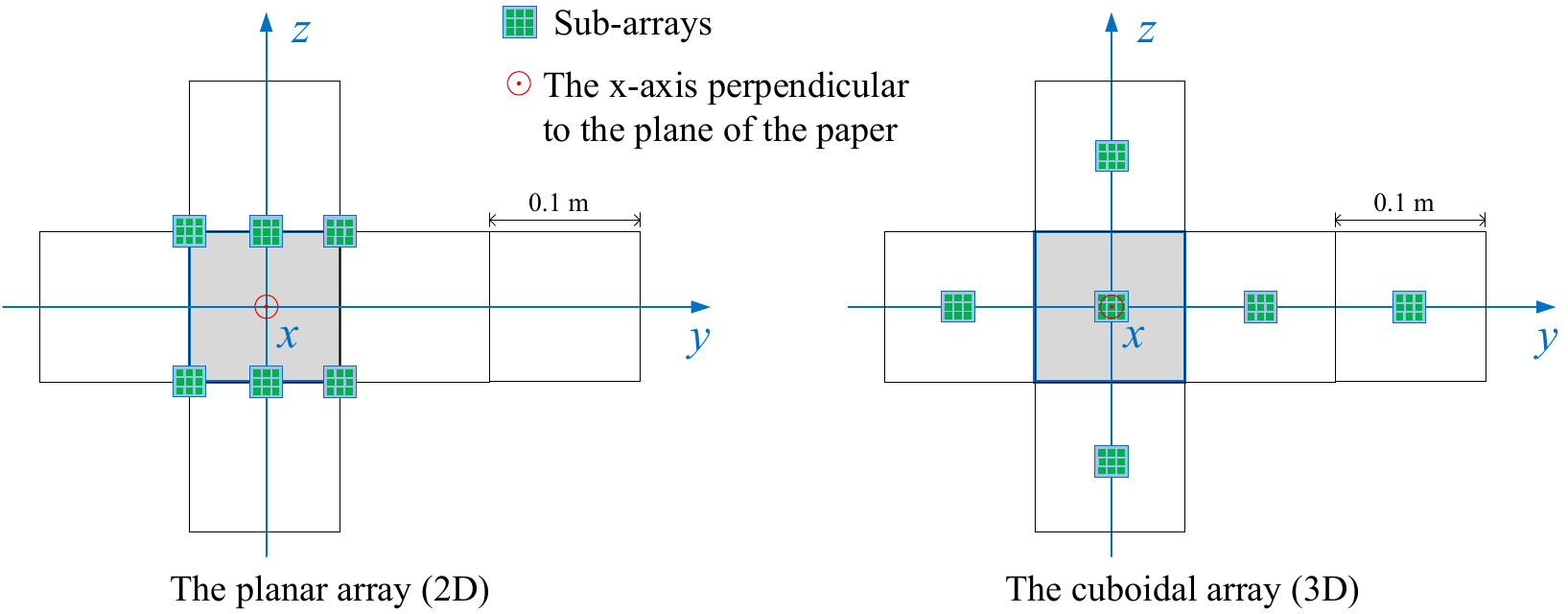}
    \caption{Illustration of the planar (2D) and cuboidal (3D) array layouts by tiling the cube into a plane.}
    \label{fig_6}
 \end{figure}

We consider the BS equipped with an $8\times 8$ uniform planar array with half-wavelength spacing between elements.
From the UE side, we evaluate two different types of array configurations (2D and 3D). Each configuration has 6 SAs, and each SA has $4\times 4$ antenna elements with half-wavelength spacing. 
For the 3D array, each SA is attached at the center of a surface of a $\unit[0.1\times0.1\times0.1]{m^3}$ cube.
On the other hand, the 2D array has all the SAs placed on the same plane.
Fig.~\ref{fig_6} shows the two array layouts where the cube is tiled into a plane.
Other parameters are set as follows, average transmission power $P = \unit[0]{dBm}$, carrier frequency $f_c = \unit[140]{GHz}$, bandwidth $W = \unit[1000]{MHz}$, number of transmissions $G = 50$, number of subcarriers $K = 10$, noise PSD $N_0 = \unit[-173.855]{dBm/Hz}$ and noise figure $N_f = \unit[10]{dBm}$.

To give an intuitive characterization of the orientation, we use Euler angles $[\alpha, \beta, \gamma ]^\mathsf{T}$ to represent a rotation matrix $\mathbf{R}$.
The rotation order is important when mapping between the Euler angles and the rotation matrix.
In this paper, we use the following rotation sequence:
\begin{equation}
    \mathbf{R} = \mathbf{R}_z(\gamma)\mathbf{R}_y(\beta)\mathbf{R}_x(\alpha),
\end{equation}
where $\mathbf{R}_x(\alpha)$ denotes a rotation of $\alpha$ degree around the X-axis, and likewise for $\mathbf{R}_y(\beta)$ and $\mathbf{R}_z(\gamma)$.
The expressions of these rotation matrices can be found in, e.g.,~\cite{nazari20213d}.

Throughout the simulations, we set the normal direction of an array as the positive direction of x-axis in its local coordinate system as shown in Fig. \ref{fig_1}. 
We consider a scenario with $M = 2$ BSs located at $[-10.5, -10.5, 5]^\mathsf{T}$ and $[10.5, 10.5, 5]^\mathsf{T}$  sending downlink signals in the form of \eqref{channel} to the UE.
The orientation of these two BSs in Euler angles are $(0^\circ,90^\circ,45^\circ)$ and $(0^\circ,90^\circ,-135^\circ)$ (facing downwards), respectively.

\subsection{Results and Discussion}
\begin{figure}[t]
    \centering
    \includegraphics[width=3.5in]{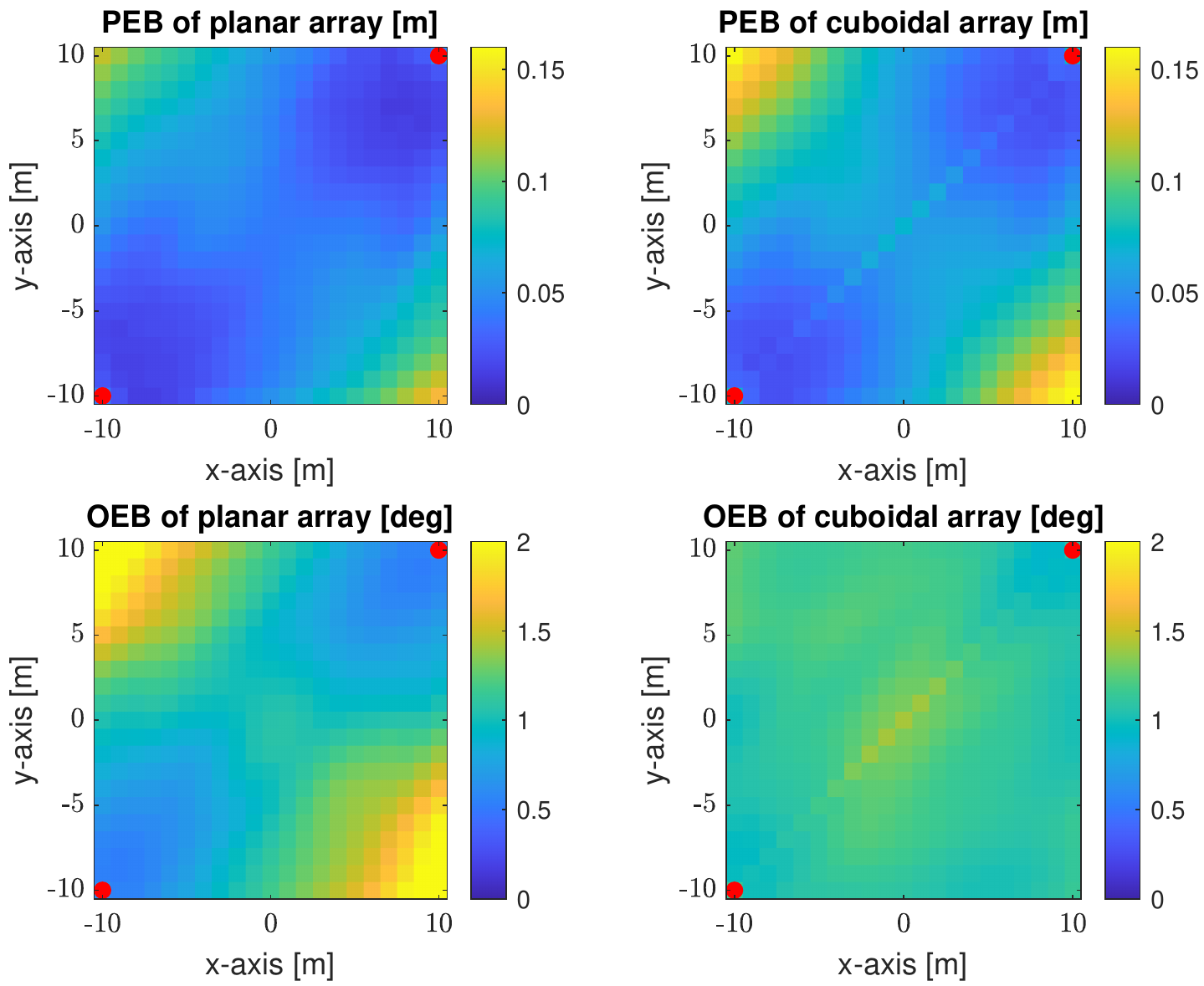}
    \caption{The PEB and OEB of the planar array and the cuboidal array across different positions. The orientation of the UE is fixed as $\alpha = 0^\circ, \beta = -90^\circ, \gamma = 45^\circ$ (facing upwards) and the height is fixed at $\unit[0]{m}$. The locations of the BSs are marked with red dots.}
    \label{fig_3}
\end{figure}

\subsubsection{PEB/OEB evaluation vs. different UE positions}
We first test the distribution of the PEB and OEB across different UE positions with a fixed orientation.
The UE orientation is set as $(0^\circ,-90^\circ,45^\circ)$ (facing upwards) and the height is fixed at $\unit[0]{m}$. The PEB/OEB is calculated in a $\unit[20\times20]{m^2}$ area with a $\unit[1]{m}$ step size as shown in Fig.~\ref{fig_3}.
We can observe that, in general, the PEB becomes larger as the UE moves away from both BSs. For this specific setup, the planar array appears to have a slightly lower PEB. This is because only a subset of the SAs of the 3D array can receive LOS signals from a BS. In contrast, all the SAs of the planar array enjoy LOS connections with the BSs.
For the OEB, we observe that the plannar array outperforms the cuboidal array when positioned between the BSs, while the 3D array has a better coverage. 
This can be explained by the fact that the OEB is largely determined by the AOA measurements. A 3D spatial arrangement can offer an advantage in AOA estimation~\cite{7096364} by maintaining LOS channel across the test area.
A final observation from Fig.~\ref{fig_3} is the higher PEB and OEB of the cuboidal array (relative to the surrounding area) in positions that fall below the (diagonal) line connecting the two BSs. In these positions, each BS can only see two of the UE SAs since the UE has a $45$-degree rotation on the horizontal plane.
In other positions, there are always three SAs that can be seen by each BS.

\begin{figure}[t]
    \centering
    \includegraphics[width=3.5in]{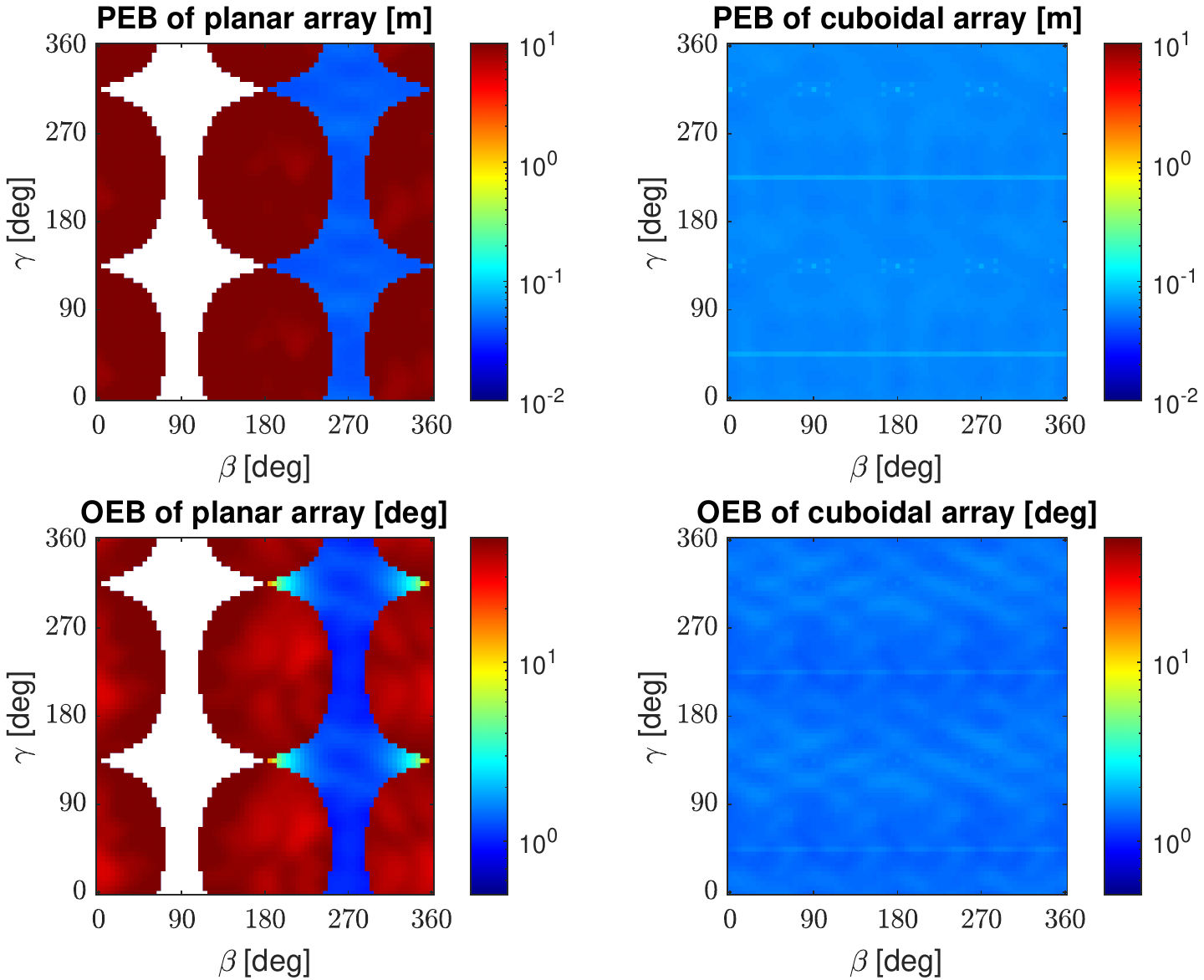}
    \caption{The PEB and OEB of the planar array and the cuboidal array across different orientations. The position of the UE is fixed at $[0,0,0]^\mathsf{T} $.}
    \label{fig_5}
\end{figure}

\subsubsection{PEB/OEB evaluation vs. different UE orientations}
In the second test, we examine the distribution of the PEB and OEB across different UE orientations for a fixed UE position at $[0,0,0]^\mathsf{T}$. We rotate the UE across $\beta,\gamma$ in the range $[0^\circ, 360^\circ]$ with the step size $5^\circ$, and fix $\alpha = 0^\circ$.
Fig.~\ref{fig_5} shows the corresponding results. 
From this figure, it is easy to conclude that, on average, the 3D array substantially outperform the 2D array in terms of localization coverage. In fact, there are only very small regions where to two configurations are comparable (e.g., around $\beta=270^\circ$). When $\beta$ is around $90^\circ$, the planar array does not have LOS to either BS; hence, both localization and communication are not possible (as indicated by the white area). 
When the UE can only establish the LOS connection with one of the two BSs, the localization is impossible without synchronization or multipath, and only communication function remains (as shown in the red area). 
On the other hand, the cuboidal array's performance is fairly consistent across all the test angles.         


\begin{figure}[htb]
\begin{minipage}[b]{0.98\linewidth}
  \centering
%
%
\definecolor{mycolor1}{rgb}{0.00000,1.00000,1.00000}%
\begin{tikzpicture}
[scale=1\columnwidth/10cm,font=\normalsize]
\begin{axis}[%
width=9cm,
height=3cm,
at={(0,0)},
scale only axis,
xmode=log,
xmin=0.001,
xmax=200,
xminorticks=true,
xticklabel style = {font=\tiny,yshift=0.5ex},
xlabel style={font=\footnotesize\color{white!15!black}, yshift=1 ex},
xlabel={PEB threshold $\xi_p$ [m]},
ymode=log,
ymin=0.001,
ymax=1,
yminorticks=true,
yticklabel style = {font=\tiny,xshift=0.5ex},
ylabel style={font=\footnotesize\color{white!15!black}, yshift= -1.5 ex},
ylabel={$1-C_p(\xi_p)$},
axis background/.style={fill=white},
xmajorgrids,
ymajorgrids,
legend style={font=\scriptsize, at={(.02, 0.02)}, anchor=south west, legend cell align=left, align=left, draw=white!5!black, legend columns=1}
]
\addplot [color=red, dashed, line width=1.0pt]
  table[row sep=crcr]{%
0.00321377485631557	1\\
0.00517543224182616	0.9993\\
0.00781542030119936	0.998\\
0.0124487691356384	0.9956\\
0.0186705669039656	0.9904\\
0.0279501778688199	0.9848\\
0.0417185901223784	0.978\\
0.062321209447348	0.9705\\
0.093707551494477	0.959800000000001\\
0.139852152561469	0.945700000000001\\
0.20883273957495	0.922700000000002\\
0.311811179961362	0.904700000000003\\
0.46683163964723	0.891100000000004\\
0.696621721634031	0.881700000000004\\
1.05117919455885	0.871400000000005\\
1.57558396258669	0.858400000000006\\
2.35262132556694	0.842800000000007\\
3.52145848398286	0.822900000000009\\
5.25955457176439	0.794800000000011\\
7.8535305894718	0.756800000000015\\
11.7210760013371	0.70530000000002\\
17.4868211046102	0.639600000000028\\
26.0893073805547	0.559000000000037\\
38.932998618345	0.45940000000004\\
58.0818149604605	0.333200000000031\\
83.1522339945544	0.223300000000024\\
112.838583380745	0.149600000000014\\
169.15443716115	0.107800000000009\\
272.684897474049	0.105000000000009\\
425.278655724301	0.104700000000009\\
648.305865406137	0.104500000000009\\
1164.31170296725	0.104300000000009\\
1607.61228191207	0.104200000000009\\
};
\addlegendentry{2 BSs (Planar)}

\addplot [color=red, line width=1.0pt]
  table[row sep=crcr]{%
0.00394519859895781	1\\
0.00591148549096145	0.9984\\
0.00883800368513104	0.9947\\
0.0132007173727278	0.9847\\
0.0198066407956929	0.966800000000001\\
0.0295695680150252	0.941400000000001\\
0.0441969867078869	0.905400000000003\\
0.0660101104768214	0.853500000000006\\
0.0985692647069242	0.775700000000013\\
0.147074994597406	0.649000000000027\\
0.219410175127795	0.464000000000041\\
0.327360329092341	0.323100000000031\\
0.475280604005696	0.216500000000023\\
0.687822867618061	0.145100000000014\\
1.02553364873781	0.0972000000000083\\
1.51608122698459	0.0651000000000058\\
2.2142035086038	0.0436000000000039\\
3.05408283025137	0.0292000000000024\\
4.25101442879755	0.0195000000000016\\
6.28562293103789	0.0130000000000011\\
8.37527667044991	0.00870000000000071\\
10.6790561635095	0.00580000000000047\\
13.2263929790069	0.00380000000000036\\
16.7638009166723	0.00250000000000017\\
21.0375904341182	0.00160000000000016\\
32.0145368513233	0.00100000000000011\\
35.504002365593	0.000600000000000045\\
37.9607947010018	0.000400000000000067\\
45.1064238958506	0.000199999999999978\\
49.3786550015562	9.9999999999989e-05\\
62.6339317785765	0\\
};
\addlegendentry{2 BSs (Cuboidal)}

\addplot [color=mycolor1, dashed, line width=1.0pt]
  table[row sep=crcr]{%
0.00224122868896719	1\\
0.00362677558045581	0.9989\\
0.00542663136118943	0.9961\\
0.00819226415972954	0.9913\\
0.0122225469865515	0.98\\
0.0182669824811898	0.969\\
0.0272822552301265	0.949700000000001\\
0.0407484026470011	0.924200000000002\\
0.0608356197849411	0.885500000000004\\
0.0908739055284702	0.837600000000007\\
0.135884801588524	0.784600000000012\\
0.202818365117429	0.732800000000017\\
0.303083586154082	0.678700000000023\\
0.452156547373415	0.62220000000003\\
0.6752382586051	0.575100000000035\\
1.00992450384714	0.545500000000038\\
1.50815871543884	0.519200000000041\\
2.25455959343592	0.499000000000043\\
3.36473292828412	0.481600000000042\\
5.06226949555169	0.46110000000004\\
7.56132571839429	0.436800000000038\\
11.298421165133	0.406900000000036\\
16.8626711454129	0.373300000000033\\
25.1608385979553	0.331600000000031\\
37.5393991994462	0.283400000000029\\
56.0812164064648	0.221000000000024\\
83.7765802492176	0.161100000000016\\
125.453257648085	0.11850000000001\\
192.797686056502	0.105400000000009\\
295.052694722384	0.104600000000009\\
648.305865406137	0.104400000000009\\
1164.31170296725	0.104200000000009\\
1164.31170296725	0.104200000000009\\
};
\addlegendentry{3 BSs (Planar)}

\addplot [color=mycolor1, line width=1.0pt]
  table[row sep=crcr]{%
0.00224373808957186	1\\
0.00347916047567224	0.9986\\
0.00520558495488861	0.9962\\
0.00784040936270329	0.991\\
0.011753164233544	0.9755\\
0.0175746375137446	0.946800000000001\\
0.0262190615901001	0.909200000000003\\
0.0391580252135708	0.849400000000006\\
0.0584665851991121	0.746900000000016\\
0.0872261868288971	0.541300000000039\\
0.124252167644244	0.362800000000032\\
0.162828645425025	0.243100000000027\\
0.209255463960246	0.162900000000016\\
0.259793274537583	0.109100000000009\\
0.308561269466112	0.0731000000000064\\
0.363346860006002	0.0490000000000044\\
0.418619248782518	0.0328000000000028\\
0.468732474855943	0.0219000000000018\\
0.51823614006173	0.0146000000000012\\
0.567563018193133	0.00970000000000082\\
0.626617878009936	0.00650000000000051\\
0.681233649915063	0.00430000000000041\\
0.705945496485931	0.00280000000000025\\
0.732395557092933	0.00180000000000013\\
0.765499703331553	0.00120000000000009\\
0.789319083689008	0.000800000000000023\\
0.812490330983125	0.000500000000000056\\
0.847407790462	0.000300000000000078\\
0.860331596063971	0.000199999999999978\\
0.877816740553667	9.9999999999989e-05\\
0.878020827721896	0\\
};
\addlegendentry{3 BSs (Cuboidal)}

\addplot [color=blue, dashdotted, line width=1.0pt]
  table[row sep=crcr]{%
0.00226598406059725	1\\
0.00346644074995861	0.9986\\
0.00536023690436872	0.9943\\
0.00803489033883614	0.9852\\
0.0119952251932998	0.9707\\
0.0178974714335381	0.948600000000001\\
0.0267152999139203	0.920200000000002\\
0.039892796773114	0.878500000000004\\
0.0596652749102274	0.817000000000009\\
0.0890230778198991	0.727500000000018\\
0.132844227163463	0.640300000000028\\
0.198189569590311	0.557100000000037\\
0.295883406304848	0.467800000000041\\
0.441415533831599	0.375700000000033\\
0.6589621217252	0.289500000000029\\
0.983753547950755	0.228200000000025\\
1.46851536335774	0.18900000000002\\
2.19090552041143	0.162400000000016\\
3.27964406967947	0.144200000000013\\
4.90415811600568	0.133200000000012\\
7.3804881187431	0.124200000000011\\
11.0184833931092	0.11760000000001\\
16.4758815805257	0.11430000000001\\
24.6426076265896	0.110900000000009\\
37.3538015900475	0.109200000000009\\
63.911805741808	0.107700000000009\\
118.247710007735	0.107400000000009\\
397.418680629925	0.107300000000009\\
397.418680629925	0.107300000000009\\
};
\addlegendentry{4 BSs (Planar)}

\addplot [color=blue, line width=1.0pt]
  table[row sep=crcr]{%
0.00203190853962711	1\\
0.00306921807879053	0.9994\\
0.00460299288262007	0.9954\\
0.00687630174873875	0.9878\\
0.0102624329471492	0.9709\\
0.0153126501708556	0.935600000000001\\
0.0228685675875398	0.883500000000004\\
0.0341305457610686	0.809000000000009\\
0.0509203763918155	0.670200000000024\\
0.0670182252224179	0.449200000000039\\
0.0770539071971637	0.301100000000029\\
0.0858918481919814	0.201800000000022\\
0.0918994870832602	0.135200000000012\\
0.0964803495330317	0.0906000000000078\\
0.0999415113268414	0.0607000000000054\\
0.102832297793057	0.0406000000000035\\
0.104786829983194	0.0272000000000023\\
0.106598766214251	0.0182000000000015\\
0.108082290436103	0.012100000000001\\
0.109263887583457	0.00810000000000066\\
0.110897289666588	0.00540000000000052\\
0.112834814346186	0.00360000000000027\\
0.113700825837012	0.00240000000000018\\
0.116703734386609	0.00160000000000016\\
0.12124962392021	0.00100000000000011\\
0.12406332873144	0.000600000000000045\\
0.124662532147836	0.000400000000000067\\
0.131851752698005	0.000199999999999978\\
0.132085552903972	9.9999999999989e-05\\
0.135269190286375	0\\
};
\addlegendentry{4 BSs (Cuboidal)}

\end{axis}
\end{tikzpicture}%
    \vspace{-0.8cm}
\end{minipage}
\hfill
\begin{minipage}[b]{0.98\linewidth}
  \centering
%
%
\definecolor{mycolor1}{rgb}{0.00000,1.00000,1.00000}%
\begin{tikzpicture}
[scale=1\columnwidth/10cm,font=\normalsize]
\begin{axis}[%
width=9cm,
height=3cm,
at={(0,0)},
scale only axis,
xmode=log,
xmin=0.0005,
xmax=50,
xminorticks=true,
xticklabel style = {font=\tiny,yshift=0.5ex},
xlabel style={font=\footnotesize\color{white!15!black}, yshift=1 ex},
xlabel={OEB threshold $\xi_o$ [deg]},
ymode=log,
ymin=0.001,
ymax=1,
yminorticks=true,
yticklabel style = {font=\tiny,xshift=0.5ex},
ylabel style={font=\footnotesize\color{white!15!black}, yshift= -1.5 ex},
ylabel={$1-C_o(\xi_o)$},
axis background/.style={fill=white},
xmajorgrids,
ymajorgrids,
legend style={font=\scriptsize, at={(.02, 0.02)}, anchor=south west, legend cell align=left, align=left, draw=white!5!black, legend columns=1}
]
\addplot [color=red, dashed, line width=1.0pt]
  table[row sep=crcr]{%
0.00988787744279652	1\\
0.0185295347239062	0.9997\\
0.0276591561985474	0.9932\\
0.0412688629656109	0.9675\\
0.0616765415399533	0.931900000000002\\
0.0923384347062284	0.906900000000003\\
0.13808540797836	0.891700000000004\\
0.206665210526993	0.882200000000004\\
0.309412821709501	0.869300000000005\\
0.461906374493165	0.850800000000006\\
0.689228783624722	0.823000000000009\\
1.03002654343277	0.779800000000012\\
1.53675329895479	0.71150000000002\\
2.29292420564922	0.62000000000003\\
3.42141112372871	0.485400000000042\\
5.10709269179051	0.334400000000031\\
7.62095660845797	0.235800000000026\\
11.3856855225076	0.179100000000019\\
16.9902396070313	0.146100000000014\\
25.5563276237848	0.127700000000011\\
38.3092748692102	0.11770000000001\\
57.3257091584356	0.11380000000001\\
85.882217195609	0.111800000000009\\
130.732455031087	0.110800000000009\\
222.566101298458	0.110400000000009\\
356.859867829975	0.109700000000009\\
532.701598173601	0.109000000000009\\
876.916138847247	0.108300000000009\\
1371.30436828515	0.107100000000009\\
2096.16390093305	0.105800000000009\\
3765.38016074861	0.105300000000009\\
6512.42135994069	0.104700000000009\\
9512.67624410014	0.104200000000009\\
};
\addlegendentry{2 BSs (Planar)}

\addplot [color=red, line width=1.0pt]
  table[row sep=crcr]{%
0.0208274477548481	1\\
0.032493592311319	0.9995\\
0.0484838936343419	0.892200000000004\\
0.0619416141552267	0.598000000000032\\
0.0815062144110277	0.400800000000035\\
0.112166151741289	0.268600000000028\\
0.161950166564783	0.180000000000019\\
0.235176992165153	0.12060000000001\\
0.346305168262226	0.0808000000000071\\
0.499467491554229	0.0541000000000048\\
0.699098773048834	0.0362000000000031\\
1.00888696057339	0.024200000000002\\
1.38527879758983	0.0162000000000013\\
1.76659389288995	0.0108000000000009\\
2.18821791197577	0.00720000000000065\\
3.17378314328166	0.00480000000000036\\
3.84838610647981	0.00320000000000031\\
4.63076580693112	0.00210000000000021\\
4.94333992518417	0.00140000000000007\\
5.39101291754477	0.000900000000000123\\
5.92869372187185	0.000600000000000045\\
6.3731775954022	0.000400000000000067\\
6.75455454164477	0.000199999999999978\\
6.78055679256805	9.9999999999989e-05\\
7.1523249464564	0\\
};
\addlegendentry{2 BSs (Cuboidal)}

\addplot [color=mycolor1, dashed, line width=1.0pt]
  table[row sep=crcr]{%
0.00988787744279652	1\\
0.0147659769258238	0.9978\\
0.0220336839843318	0.958200000000001\\
0.0328848812586802	0.859400000000006\\
0.0490605146392865	0.749000000000016\\
0.0732437805037895	0.657200000000026\\
0.109513830541538	0.598400000000032\\
0.163421939630106	0.559100000000037\\
0.243848648061209	0.534800000000039\\
0.365049216435302	0.509700000000042\\
0.545725749292337	0.487900000000042\\
0.8143472331335	0.463700000000041\\
1.21561860667267	0.431500000000038\\
1.81406904827302	0.387600000000034\\
2.70640325416786	0.331600000000031\\
4.03841119370907	0.259800000000028\\
6.02683897852692	0.195700000000021\\
9.01053942568293	0.158000000000015\\
13.4727787253908	0.134400000000012\\
20.121121694598	0.12260000000001\\
30.1509076119885	0.11440000000001\\
45.5684987632414	0.110000000000009\\
68.5100627049083	0.108900000000009\\
104.48318477729	0.108100000000009\\
259.035307503632	0.107600000000009\\
409.892767266945	0.107100000000009\\
678.947619045273	0.106700000000009\\
1017.92313633972	0.106000000000009\\
1522.98346931693	0.105700000000009\\
4320.1127114385	0.105000000000009\\
6512.42135994069	0.104700000000009\\
9512.67624410014	0.104200000000009\\
};
\addlegendentry{3 BSs (Planar)}

\addplot [color=mycolor1, line width=1.0pt]
  table[row sep=crcr]{%
0.0124894173642895	1\\
0.0191153904492414	0.9992\\
0.0280236212267777	0.669700000000024\\
0.0320612628600523	0.448900000000039\\
0.0361262971854822	0.300900000000029\\
0.040020652411007	0.201600000000022\\
0.0437359410920901	0.135100000000012\\
0.047298663040945	0.0905000000000078\\
0.0512767848323483	0.0606000000000054\\
0.0550236261307899	0.0406000000000035\\
0.0591254274768609	0.0272000000000023\\
0.0624018822887527	0.0182000000000015\\
0.0656692138947683	0.012100000000001\\
0.0687157150345736	0.00810000000000066\\
0.0719508195419081	0.00540000000000052\\
0.074757559326625	0.00360000000000027\\
0.0763385074122478	0.00240000000000018\\
0.0779609308104053	0.00160000000000016\\
0.0797930482355357	0.00100000000000011\\
0.0855268870973913	0.000600000000000045\\
0.0871855691864493	0.000400000000000067\\
0.0872069966504657	0.000199999999999978\\
0.088612366940436	9.9999999999989e-05\\
0.0913141758777622	0\\
};
\addlegendentry{3 BSs (Cuboidal)}

\addplot [color=blue, dashdotted, line width=1.0pt]
  table[row sep=crcr]{%
0.00900010743584662	1\\
0.013561310690623	0.9963\\
0.0202435643439805	0.945900000000001\\
0.0302160579408396	0.790500000000011\\
0.0450807038820125	0.604700000000032\\
0.0672771949072797	0.438100000000038\\
0.100407892348784	0.329400000000031\\
0.1499097151116	0.253100000000028\\
0.223665490225824	0.202300000000022\\
0.333806853578296	0.169200000000017\\
0.498555296954324	0.147700000000014\\
0.751198448971513	0.133200000000012\\
1.12219704249745	0.123400000000011\\
1.6894623471922	0.11750000000001\\
2.53111158794127	0.11290000000001\\
3.79967988681564	0.110100000000009\\
5.85494427629401	0.108600000000009\\
9.20399491839167	0.107800000000009\\
13.8586050511356	0.107700000000009\\
31.1228958372541	0.107400000000009\\
53.147851890805	0.107300000000009\\
53.147851890805	0.107300000000009\\
};
\addlegendentry{4 BSs (Planar)}

\addplot [color=blue, line width=1.0pt]
  table[row sep=crcr]{%
0.0115416281414167	1\\
0.0173674398569388	0.9993\\
0.0224628799202051	0.669800000000024\\
0.0232622990309393	0.448900000000039\\
0.0237894761028864	0.300900000000029\\
0.0242407637428908	0.201600000000022\\
0.0247195986168492	0.135100000000012\\
0.0253089601323678	0.0905000000000078\\
0.0259254684603623	0.0606000000000054\\
0.0265070521978093	0.0406000000000035\\
0.0270881864899894	0.0272000000000023\\
0.0275925554291666	0.0182000000000015\\
0.0282168883957473	0.012100000000001\\
0.0287187740680756	0.00810000000000066\\
0.0291721070451533	0.00540000000000052\\
0.029682807829483	0.00360000000000027\\
0.0303562421590723	0.00240000000000018\\
0.0308081562099251	0.00160000000000016\\
0.031522618485242	0.00100000000000011\\
0.0322200474973032	0.000600000000000045\\
0.0325912180708779	0.000400000000000067\\
0.0340906914933193	0.000199999999999978\\
0.0352395995062225	9.9999999999989e-05\\
0.0368157402029504	0\\
};
\addlegendentry{4 BSs (Cuboidal)}

\end{axis}
\end{tikzpicture}%
    \vspace{-0.8cm}
\end{minipage}
\caption{Empirical CCDF of the PEB (top) and OEB (bottom) of the planar array and the cuboidal array under different number of BSs.}
\label{fig_4}
\end{figure}
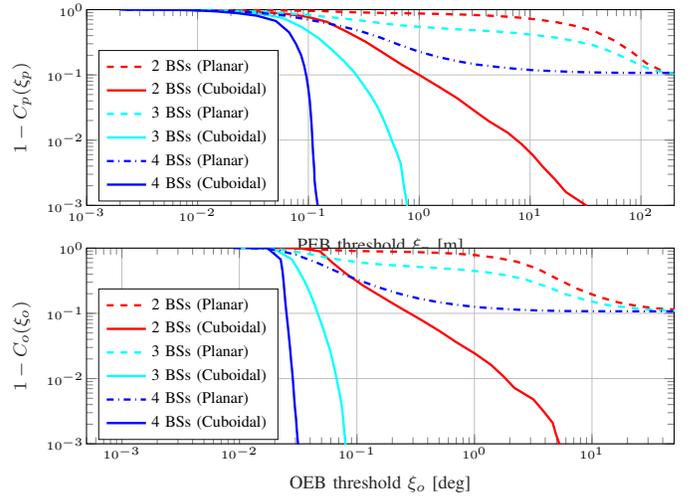

\subsubsection{Coverage evaluation}
Finally, we test the localization coverage of the 2D and 3D array configurations defined in~\eqref{eq_defcoverage}.
The UE's position and orientation are uniform distributed; namely, $x,y\sim\text{U}(-10,10)$, $z\sim\text{U}(0,5)$, and $\alpha,\beta,\gamma\sim \text{U}(0,360)$.
To give a compact view of the PEB/OEB's threshold with the coverage in different order of magnitude $\{90\%,99\%,99.9\%,...\}$ (i.e., the outage in different order of magnitude $\{10\%,1\%,0.1\%,...\}$), we demonstrate $1-C_p$ and $1-C_o$ over different threshold $\xi_p$ and $\xi_o$, which is empirical complementary cumulative distribution function (CCDF). 
We test the CCDF for the cases there are $M = \{2,3,4\}$ BSs in the system.
For 3 BSs case, we add one BS at location $[-10.5,10.5,5]^\mathsf{T}$ with orientation $(0^\circ,90^\circ,-45^\circ)$; For 4 BSs case, we add one more BS at location $[10.5,-10.5,5]^\mathsf{T}$ with orientation $(0^\circ,90^\circ,135^\circ)$.
We use 10000 simulation trials to compute the PEB and OEB and plot the CCDF curve, as shown in Fig.~\ref{fig_4}.
We observe that the planar array suffers an outage in $10\%$ of the cases. As explained earlier, this is due to the lack of LOS with all the BSs. 
We can also see that, as an example, for the 4 BSs case and coverage of $90\%$, we have a PEB within about $0.1$~m using the cuboidal array, while the planar array gives a PEB within more than $10$~m. 
This reveals that the cuboidal (3D) array is able to achieve better coverage than the planar (2D) array. Generally, under the same threshold, the more BS we deploy, the lower the outage and thus the higher the coverage we can obtain for both 2D and 3D arrays.

\section{Conclusion}

In this paper, we studied the far-field localization problem of a UE equipped with a 3D array in a THz-band downlink system  with multiple \acp{bs}. We derived the PEB and OEB based on the Constrained Cram\'{e}r–Rao bound, and defined the localization coverage to evaluate the performance of of 2D and 3D array configurations. Based on extensive simulations across different UE positions and orientations, we found that the cuboidal array can achieve better coverage, while the planar array has a lower error bound in certain positions and orientations. 
This work is instructive for the BS placement optimization and array design of the THz localization systems, which can be potential future research directions.

\section*{Acknowledgment}
This publication is based upon work supported by the King Abdullah University of Science and Technology (KAUST) Office of Sponsored Research (OSR) under Award No. ORA-CRG2021-4695, 
and by the European Commission through the H2020 project Hexa-X (Grant Agreement No. 101015956).

\appendices

\section{The expression of Transformation matrix}\label{appendix_1}

Let 
\footnotesize
\begin{align}\nonumber
  &\mathbf{v} =  \mathbf{p}_{\text{U}}+\mathbf{R}_{\text{U}}\mathbf{d}_n-\mathbf{p}_{\text{B},m},
\\ \nonumber
    &s_1 = \left[1-\left(\frac{ \mathbf{u}_3^\mathsf{T} \mathbf{R}_{\text{B},m}^\mathsf{T} \mathbf{v} }{ \|\mathbf{v}\|_2 }\right)^2\right]^{-\frac{1}{2}},\ 
    s_2 = \left[1-\left(\frac{ \mathbf{u}_3^\mathsf{T} \mathbf{R}_n^\mathsf{T} \mathbf{R}_{\text{U}}^\mathsf{T} \mathbf{v} }{ \|\mathbf{v}\|_2 }\right)^2\right]^{-\frac{1}{2}},   
\\ \nonumber
    &\mathbf{K}_1 = (\mathbf{p}_{\text{U}}-\mathbf{p}_{\text{B},m})\mathbf{u}_2^\mathsf{T} \mathbf{R}_n^\mathsf{T}  + \mathbf{R}_{\text{U}}(\mathbf{R}_n\mathbf{u}_2\mathbf{d}_n^\mathsf{T}  + \mathbf{d}_n\mathbf{u}_2^\mathsf{T} \mathbf{R}_n^\mathsf{T} ),   
\\ \nonumber
    &\mathbf{K}_2 = (\mathbf{p}_{\text{U}}-\mathbf{p}_{\text{B},m})\mathbf{u}_1^\mathsf{T} \mathbf{R}_n^\mathsf{T}  + \mathbf{R}_{\text{U}}(\mathbf{R}_n\mathbf{u}_1\mathbf{d}_n^\mathsf{T}  + \mathbf{d}_n\mathbf{u}_1^\mathsf{T} \mathbf{R}_n^\mathsf{T} ),   
\\ \nonumber
    &\mathbf{g} = \text{vec}\left(
    (\mathbf{p}_{\text{U}}-\mathbf{p}_{\text{B},m})\mathbf{u}_3^\mathsf{T} \mathbf{R}_n^\mathsf{T}  + \mathbf{R}_{\text{U}}(\mathbf{R}_n\mathbf{u}_3\mathbf{d}_n^\mathsf{T} +\mathbf{d}_n\mathbf{u}_3^\mathsf{T} \mathbf{R}_n^\mathsf{T} )
  \right).
\end{align}
\normalsize
Then according to \eqref{eq_forward1}--\eqref{eq_forward5}, we can derive:
\footnotesize
\begin{align}\nonumber
&\dfrac{\partial\theta_{m,n}^{(\text{az})}}{\partial \mathbf{p}_{\text{U}}} =
\frac{\left(\mathbf{u}_1^\mathsf{T} \mathbf{R}_{\text{B},m}^\mathsf{T} \mathbf{v}\right) \mathbf{R}_{\text{B},m}\mathbf{u}_2 - 
\left(\mathbf{u}_2^\mathsf{T} \mathbf{R}_{\text{B},m}^\mathsf{T} \mathbf{v}\right) \mathbf{R}_{\text{B},m}\mathbf{u}_1}
{\left(\mathbf{u}_1^\mathsf{T} \mathbf{R}_{\text{B},m}^\mathsf{T} \mathbf{v}\right)^2+\left(\mathbf{u}_2^\mathsf{T} \mathbf{R}_{\text{B},m}^\mathsf{T} \mathbf{v}\right)^2},
\\ \nonumber
&\dfrac{\partial\theta_{m,n}^{(\text{az})}}{\partial \mathbf{R}_{\text{U}}} =
\frac{\left(\mathbf{u}_1^\mathsf{T} \mathbf{R}_{\text{B},m}^\mathsf{T} \mathbf{v}\right) \mathbf{R}_{\text{B},m}\mathbf{u}_2 - 
\left(\mathbf{u}_2^\mathsf{T} \mathbf{R}_{\text{B},m}^\mathsf{T} \mathbf{v}\right) \mathbf{R}_{\text{B},m}\mathbf{u}_1}
{\left(\mathbf{u}_1^\mathsf{T} \mathbf{R}_{\text{B},m}^\mathsf{T} \mathbf{v}\right)^2+\left(\mathbf{u}_2^\mathsf{T} \mathbf{R}_{\text{B},m}^\mathsf{T} \mathbf{v}\right)^2}\mathbf{d}_n^\mathsf{T} ,
\\ \nonumber
&\dfrac{\partial\theta_{m,n}^{(\text{el})}}{\partial \mathbf{p}_{\text{U}}} =
s_1\cdot\left[\frac{\mathbf{R}_{\text{B},m}\cdot\mathbf{u}_3}{\|\mathbf{v}\|_2} - 
\frac{\left(\mathbf{u}_3^\mathsf{T} \mathbf{R}_{\text{B},m}^\mathsf{T} \mathbf{v}\right)\mathbf{v}}{ \|\mathbf{v}\|_2^3 }\right],
\\ \nonumber
&\text{vec}\left(\dfrac{\partial\theta_{m,n}^{(\text{el})}}{\partial \mathbf{R}_{\text{U}}}\right) = s_1
\Biggl[\frac{\text{vec}(\mathbf{R}_{\text{B},m}\mathbf{u}_3\mathbf{d}_n^\mathsf{T} )}{\|\mathbf{v}\|_2}-
\frac{\mathbf{u}_3^\mathsf{T} \mathbf{R}_{\text{B},m}^\mathsf{T} \mathbf{v}}{\|\mathbf{v}\|_2^3}
\begin{bmatrix}
  \mathbf{d}_n(1)\mathbf{I}_3\\
  \mathbf{d}_n(2)\mathbf{I}_3\\
  \mathbf{d}_n(3)\mathbf{I}_3\\
\end{bmatrix}
\mathbf{v}\Biggl],
\\ \nonumber
&\dfrac{\partial\phi_{m,n}^{(\text{az})}}{\partial \mathbf{p}_{\text{U}}} = 
\frac{\left(\mathbf{u}_1^\mathsf{T} \mathbf{R}_n^\mathsf{T} \mathbf{R}_{\text{U}}^\mathsf{T} \mathbf{v}\right)\mathbf{R}_{\text{U}}\mathbf{R}_{n}\mathbf{u}_2
- \left(\mathbf{u}_2^\mathsf{T} \mathbf{R}_n^\mathsf{T} \mathbf{R}_{\text{U}}^\mathsf{T} \mathbf{v}\right)\mathbf{R}_{\text{U}}\mathbf{R}_{n}\mathbf{u}_1  }
{\left(\mathbf{u}_1^\mathsf{T} \mathbf{R}_n^\mathsf{T} \mathbf{R}_{\text{U}}^\mathsf{T} \mathbf{v}\right)^2 + \left(\mathbf{u}_2^\mathsf{T} \mathbf{R}_n^\mathsf{T} \mathbf{R}_{\text{U}}^\mathsf{T} \mathbf{v}\right)^2},
\\ \nonumber
&\dfrac{\partial\phi_{m,n}^{(\text{az})}}{\partial \mathbf{R}_{\text{U}}} = 
\frac{\left(\mathbf{u}_1^\mathsf{T} \mathbf{R}_n^\mathsf{T} \mathbf{R}_\text{U}^\mathsf{T}\mathbf{v}\right)\cdot \mathbf{K}_1
- \left(\mathbf{u}_2^\mathsf{T} \mathbf{R}_n^\mathsf{T} \mathbf{R}_\text{U}^\mathsf{T} \mathbf{v}\right)\cdot \mathbf{K}_2 }
{\left(\mathbf{u}_1^\mathsf{T} \mathbf{R}_n^\mathsf{T} \mathbf{R}_{\text{U}}^\mathsf{T} \mathbf{v}\right)^2 + \left(\mathbf{u}_2^\mathsf{T} \mathbf{R}_n^\mathsf{T} \mathbf{R}_{\text{U}}^\mathsf{T} \mathbf{v}\right)^2},
\\ \nonumber
&\dfrac{\partial\phi_{m,n}^{(\text{el})}}{\partial \mathbf{p}_{\text{U}}} = -s_2\cdot\left[
    \frac{\mathbf{R}_{\text{U}}\mathbf{R}_n\mathbf{u}_3}{\|\mathbf{v}\|_2}-
    \frac{\mathbf{u}_3^\mathsf{T} \mathbf{R}_n^\mathsf{T} \mathbf{R}_{\text{U}}^\mathsf{T} \mathbf{v}}{\|\mathbf{v}\|_2^3}\mathbf{v}
  \right],
\\ \nonumber
    &\text{vec}\left(\dfrac{\partial\phi_{m,n}^{(\text{el})}}{\partial \mathbf{R}_{\text{U}}}\right) = -s_2\cdot
    \frac{\mathbf{g}
    }{\|\mathbf{v}\|_2}+
    s_2\cdot\frac{\mathbf{u}_3^\mathsf{T} \mathbf{R}_n^\mathsf{T} \mathbf{R}_\text{U}^\mathsf{T} \mathbf{v}}{\|\mathbf{v}\|_2^3
    }
    \begin{bmatrix}
      \mathbf{d}_n(1)\mathbf{I}_3\\
      \mathbf{d}_n(2)\mathbf{I}_3\\
      \mathbf{d}_n(3)\mathbf{I}_3\\
    \end{bmatrix}
    \mathbf{v},
\\ \nonumber
  &\dfrac{\partial\tau_{m,n}}{\partial \mathbf{p}_{\text{U}}} = \frac{1}{c}\cdot\frac{
    \mathbf{v}
  }
  {\|\mathbf{v}\|_2},\ \  \dfrac{\partial\tau_{m,n}}{\partial \rho} = 1,
\\ \nonumber
  &\text{vec}\left(\dfrac{\partial\tau_{m,n}}{\partial \mathbf{R}_{\text{U}}}\right) = 
  \frac{1}{c}\cdot
  \begin{bmatrix}
    \mathbf{d}_n(1)\mathbf{I}_3\\
    \mathbf{d}_n(2)\mathbf{I}_3\\
    \mathbf{d}_n(3)\mathbf{I}_3\\
  \end{bmatrix}\cdot
  \frac{
    \mathbf{v}
  }
  {\|\mathbf{v}\|_2}.
  \end{align}

\normalsize

\bibliography{main_0413}

\begin{thebibliography}{10}
\providecommand{\url}[1]{#1}
\csname url@samestyle\endcsname
\providecommand{\newblock}{\relax}
\providecommand{\bibinfo}[2]{#2}
\providecommand{\BIBentrySTDinterwordspacing}{\spaceskip=0pt\relax}
\providecommand{\BIBentryALTinterwordstretchfactor}{4}
\providecommand{\BIBentryALTinterwordspacing}{\spaceskip=\fontdimen2\font plus
\BIBentryALTinterwordstretchfactor\fontdimen3\font minus
  \fontdimen4\font\relax}
\providecommand{\BIBforeignlanguage}[2]{{%
\expandafter\ifx\csname l@#1\endcsname\relax
\typeout{** WARNING: IEEEtran.bst: No hyphenation pattern has been}%
\typeout{** loaded for the language `#1'. Using the pattern for}%
\typeout{** the default language instead.}%
\else
\language=\csname l@#1\endcsname
\fi
#2}}
\providecommand{\BIBdecl}{\relax}
\BIBdecl

\bibitem{9112745}
H.~Sarieddeen, N.~Saeed, T.~Y. Al-Naffouri, and M.-S. Alouini, ``Next
  generation terahertz communications: A rendezvous of sensing, imaging, and
  localization,'' \emph{IEEE Commun. Mag.}, vol.~58, no.~5, pp. 69--75, 2020.

\bibitem{sarieddeen2021overview}
H.~Sarieddeen, M.-S. Alouini, and T.~Y. Al-Naffouri, ``An overview of signal
  processing techniques for terahertz communications,'' \emph{Proc. IEEE}, vol.
  109, no.~10, pp. 1628--1665, 2021.

\bibitem{8240645}
A.~Shahmansoori, G.~E. Garcia, G.~Destino, G.~Seco-Granados, and H.~Wymeersch,
  ``Position and orientation estimation through millimeter-wave {MIMO} in {5G}
  systems,'' \emph{IEEE Trans. Wireless Commun.}, vol.~17, no.~3, pp.
  1822--1835, 2018.

\bibitem{9721709}
S.~Wang, X.~Jiang, and H.~Wymeersch, ``Cooperative localization in wireless
  sensor networks with {AOA} measurements,'' \emph{IEEE Trans. Wireless
  Commun.}, early access. doi: 10.1109/TWC.2022.3152426.

\bibitem{saeedi2018navigating}
S.~Saeedi, B.~Bodin, H.~Wagstaff, A.~Nisbet, L.~Nardi, J.~Mawer, N.~Melot,
  O.~Palomar, E.~Vespa, T.~Spink \emph{et~al.}, ``Navigating the landscape for
  real-time localization and mapping for robotics and virtual and augmented
  reality,'' \emph{Proc. IEEE}, vol. 106, no.~11, pp. 2020--2039, 2018.

\bibitem{9665433}
S.~Bartoletti, H.~Wymeersch, T.~Mach, O.~Brunnegård, D.~Giustiniano,
  P.~Hammarberg, M.~F. Keskin, J.~O. Lacruz, S.~M. Razavi, J.~Rönnblom,
  F.~Tufvesson, J.~Widmer, and N.~B. Melazzi, ``Positioning and sensing for
  vehicular safety applications in {5G} and beyond,'' \emph{IEEE Commun. Mag.},
  vol.~59, no.~11, pp. 15--21, 2021.

\bibitem{liu2021constrained}
X.~Liu, T.~Ballal, H.~Chen, and T.~Y. Al-Naffouri, ``Constrained wrapped least
  squares: {A} tool for high accuracy {GNSS} attitude determination,'' 2021,
  arXiv: 2112.14813.

\bibitem{chen2021tutorial}
H.~Chen, H.~Sarieddeen, T.~Ballal, H.~Wymeersch, M.-S. Alouini, and T.~Y.
  Al-Naffouri, ``A tutorial on terahertz-band localization for {6G}
  communication systems,'' \emph{Accepted for publication in {IEEE} Commun.
  Surveys Tuts. arXiv preprint arXiv:2110.08581}, 2022.

\bibitem{nazari20213d}
M.~A. Nazari, G.~Seco-Granados, P.~Johannisson, and H.~Wymeersch, ``{3D}
  orientation estimation with multiple {5G} mmwave base stations,'' in
  \emph{Proc. IEEE Int. Conf. Commun. (ICC)}, 2021.

\bibitem{6784422}
N.~Guzey, H.~Xu, and S.~Jagannathan, ``Localization of near-field radio
  controlled unintended emitting sources in the presence of multipath fading,''
  \emph{IEEE Trans. Instrum. Meas.}, vol.~63, no.~11, pp. 2696--2703, 2014.

\bibitem{9086740}
J.~Zou, Y.~Sun, and Q.~Wan, ``A novel {3-D} localization scheme using {1-D}
  angle measurements,'' \emph{IEEE Sens. Lett.}, vol.~4, no.~6, pp. 1--4, 2020.

\bibitem{6850012}
L.~Kumar, A.~Tripathy, and R.~M. Hegde, ``Robust multi-source localization over
  planar arrays using {MUSIC}-group delay spectrum,'' \emph{IEEE Trans. Signal
  Process.}, vol.~62, no.~17, pp. 4627--4636, 2014.

\bibitem{8058460}
T.~Pavlenko, C.~Reustle, Y.~Dobrev, M.~Gottinger, L.~Jassoume, and M.~Vossiek,
  ``Design and optimization of sparse planar antenna arrays for wireless {3-D}
  local positioning systems,'' \emph{IEEE Trans. Antennas Propag.}, vol.~65,
  no.~12, pp. 7288--7297, 2017.

\bibitem{6638438}
A.~Hassanien, S.~A. Vorobyov, and J.-Y. Park, ``Joint transmit array
  interpolation and transmit beamforming for source localization in {MIMO}
  radar with arbitrary arrays,'' in \emph{Proc. IEEE Int. Conf. Acoustics,
  Speech, and Signal Process. (ICASSP)}, 2013, pp. 4139--4143.

\bibitem{8474363}
Z.~Peng and C.~Li, ``A portable {$ K $}-band {3-D MIMO} radar with nonuniformly
  spaced array for short-range localization,'' \emph{IEEE Trans. Microwave
  Theory Tech.}, vol.~66, no.~11, pp. 5075--5086, 2018.

\bibitem{7740042}
F.~Guidi, A.~Guerra, D.~Dardari, A.~Clemente, and R.~D’Errico, ``Joint energy
  detection and massive array design for localization and mapping,'' \emph{IEEE
  Trans. Wireless Commun.}, vol.~16, no.~3, pp. 1359--1371, 2017.

\bibitem{albanese2022loko}
A.~Albanese, V.~Sciancalepore, A.~Banchs, and X.~Costa-P{\'e}rez, ``{LOKO}:
  Localization-aware roll-out planning for future mobile networks,'' 2022,
  arXiv: 2201.04051.

\bibitem{7096364}
D.~T. Vu, A.~Renaux, R.~Boyer, and S.~Marcos, ``Performance analysis of {2D}
  and {3D} antenna arrays for source localization,'' in \emph{Proc. Eur. Signal
  Process. Conf.}, 2010, pp. 661--665.

\bibitem{lin2016terahertz}
C.~Lin and G.~Y. Li, ``Terahertz communications: An array-of-subarrays
  solution,'' \emph{IEEE Commun. Mag.}, vol.~54, no.~12, pp. 124--131, 2016.

\bibitem{9591285}
S.~Tarboush, H.~Sarieddeen, H.~Chen, M.~H. Loukil, H.~Jemaa, M.-S. Alouini, and
  T.~Y. Al-Naffouri, ``{TeraMIMO}: {A} channel simulator for wideband
  ultra-massive {MIMO} terahertz communications,'' \emph{IEEE Trans. Veh.
  Technol.}, vol.~70, no.~12, pp. 12\,325--12\,341, 2021.

\bibitem{kay1993fundamentals}
S.~M. Kay, \emph{Fundamentals of statistical signal processing: estimation
  theory}.\hskip 1em plus 0.5em minus 0.4em\relax Prentice-Hall, Inc., 1993.

\bibitem{8356190}
Z.~Abu-Shaban, X.~Zhou, T.~Abhayapala, G.~Seco-Granados, and H.~Wymeersch,
  ``Error bounds for uplink and downlink {3D} localization in {5G} millimeter
  wave systems,'' \emph{IEEE Trans. Wireless Commun.}, vol.~17, no.~8, pp.
  4939--4954, 2018.

\bibitem{stoica1998cramer}
P.~Stoica and B.~C. Ng, ``On the {C}ram{\'e}r-{R}ao bound under parametric
  constraints,'' \emph{IEEE Signal Process. Lett.}, vol.~5, no.~7, pp.
  177--179, 1998.

\bibitem{liu2011localization}
L.~Liu, X.~Zhang, and H.~Ma, ``Localization-oriented coverage in wireless
  camera sensor networks,'' \emph{IEEE Trans. Wireless Commun.}, vol.~10,
  no.~2, pp. 484--494, 2011.

\bibitem{9271904}
O.~M. Bushnaq, M.~A. Kishk, A.~Celik, M.-S. Alouini, and T.~Y. Al-Naffouri,
  ``Optimal deployment of tethered drones for maximum cellular coverage in user
  clusters,'' \emph{IEEE Trans. Wireless Commun.}, vol.~20, no.~3, pp.
  2092--2108, 2021.

\end{thebibliography}
\bibliographystyle{IEEEtran}

\end{document}